# Work function of $\alpha$–$Fe_2O_3$: a DFT calculation

Chunshan He

School of Physics, Sun Yat-sen University, Guangzhou, 510275, People's Republic of China

**Abstract**: The work functions of (001) and (00 -1) surfaces of $\alpha-Fe_2O_3$ are investigated with density functional theory and symmetry slab model. These two surfaces are found to be almost nonpolarized and their work functions are 6.10 eV and 5.49 eV respectively.

## 1. Introduction

$\alpha-Fe_2O_3$(hematite) is crystallized in the rhombohedral lattice system, and it is mined as the main ore of iron. $Fe_2O_3$ will be covered at the surface of the steel or iron when they are exposed in the oxygen or water, and we call it rust. Hematite is harder than pure iron, but much more brittle, so it will weaken the iron material's strength. Hematite is one of the transition metal oxides which are an important class of functional materials[1]. Many theoretical works study its geometry, surface and bulk electric properties and magnetic properties[2-4]. The work function(WF) of $Fe_2O_3$(001) is 5.6eV which was calculated by the point charges plus quantum cluster algorithm[5]. Chen et al. considered the strain effect and used density functional theory(DFT) to calculate hematite's work function relation with the strain[6]. They found the electronic properties of Fe and O terminated $Fe_2O_3$ thin films under strain. On the other hand, more and more steel materials are used to act as field electron emission tool, such as the wireless transmit antenna, lightning conductor, and so on. For these devices, the covered $Fe_2O_3$ slab will affect the field emission. Work function is a crucial feature of field emitter. The present paper aims to calculate the work function of $Fe_2O_3$ slab which is formed under natural conditions with DFT method.

## 2. Computational method and model

The simulation was done via the Vienna *Ab inito* Simulation Package (VASP) [7]. The electron-core interactions were treated in the projector augmented wave (PAW) approximation [8]. The density functional is treated by the local density approximation (LDA) (with the Ceperly–Alder exchange correlation potential [9]). A kinetic energy cutoff (400 eV) was used. The k point mesh 9×9×1 is tested to converge.

For the atom of Fe, the localized *d* electrons need to consider the on-site Coulomb interaction. They are usually specified in terms of the effective on-site Coulomb and exchange parameters: *U* and *J*. The simplified method(LDA+U) is to use the following form(Dudarev's approach)[10]:

$$E_{LDA+U} = E_{LDA} + \frac{U-J}{2} \sum_{\sigma} \left[ \left( \sum_{m_1} n^{\sigma}_{m_1,m_1} \right) - \left( \sum_{m_1,m_2} \hat{n}^{\sigma}_{m_1,m_2} \hat{n}^{\sigma}_{m_2,m_1} \right) \right] \quad (1)$$

In equation (1), only the difference $\Delta = (U-J)$ is meaningful. $\Delta = 5eV$ is used when the simulation is done with VASP, the calculation results are agreement with the experiments[11].

$Fe_2O_3$ belongs to the space group: R3c (No. 167), lattice constants: $a = b = 0.50259$ nm and $c = 1.3735$ nm[12]. There are 12 Iron atoms and 18 Oxygen atoms in a unit cell. We first optimized the atomic coordinates in a unit cell with period boundary condition. Then a slab model was built with a unit cell duplicated in c direction (Z direction) preserving the period boundary condition in *a* and *b* direction. Vacuum slabs each with thickness $dz = 1.0$ nm were inserted between adjacent $Fe_2O_3$ slabs in Z direction. All atoms were fully relaxed until the force on each atom is less than 0.01 eV/Å. The optimized $Fe_2O_3$ slab super cell is shown in Fig. 1.

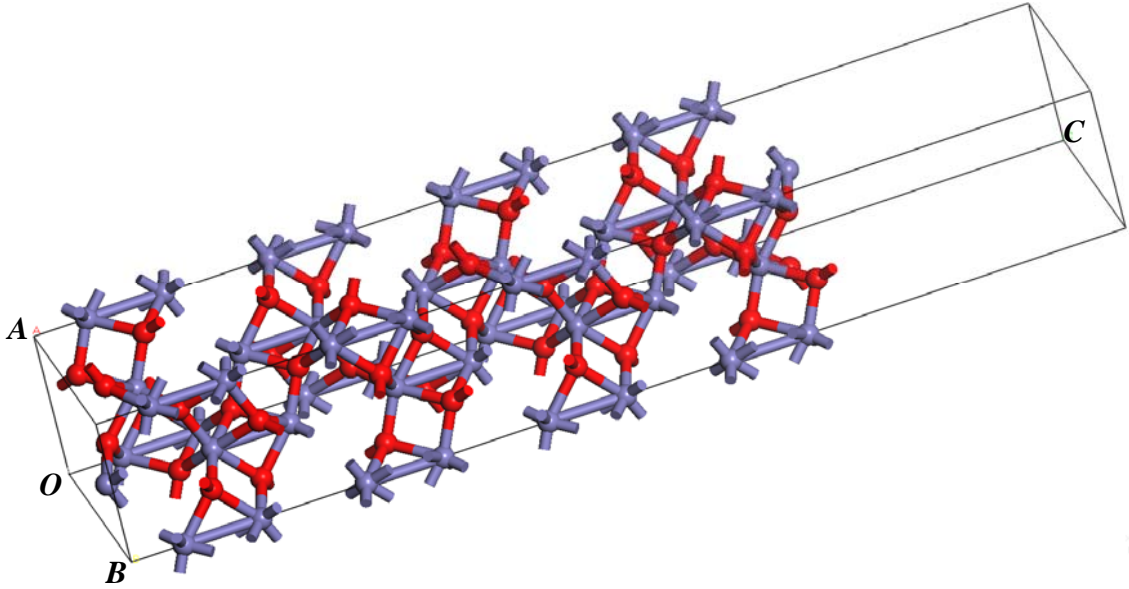

Fig. 1 The optimized $Fe_2O_3$ slab super cell formed via two unit cells of $Fe_2O_3$ and one vacuum slab. The grey balls are iron atoms and the red balls are oxygen atoms. The (00-1) surface is on the left and (001) on the right. The thicknesses of the vacuum slabs are subjected to change.

The WF is defined as WF = $\phi$ - $E_f$ with $E_f$ the Fermi level and $\phi$ the vacuum potential energy. Note that the two surfaces ((001) and (00 -1)) of the $Fe_2O_3$ slab are not symmetric. Therefore, the vacuum potential energy (and thus the WF) may not be the same outside these two surfaces, because the vacuum potential energy is modulated by the dipole layer at the surface [13-24]. Further more, two surfaces have opposite charge would produce a uniform electric field in the vacuum slab and an opposite uniform electric field in the $Fe_2O_3$ slab. Between the $Fe_2O_3$ slab and its adjacent duplicate copy one can not find a point where potential energy is constant, thus can not define the vacuum energy. To eliminate this uniform electric field in the vacuum, we use two $Fe_2O_3$ slabs as a supper cell. The atomic structures of two adjacent surfaces of two slabs are the same (Fig. 2). The potential energy at the middle of the vacuum then is constant and defined as the vacuum potential energy. There are two different vacuum regions (region II and region IV in Fig. 2) and two vacuum potential energies. When we calculate the WF of a surface of $Fe_2O_3$, the thickness of the vacuum slab of that surface should be large enough such that the potential energy in the middle of the vacuum slab is not sensitive to the thickness. We set the thickness $dz_1 = 30$ Angstrom for this vacuum slab. To save computational resource, we let the thickness of the other vacuum slab be smaller ($dz_2 = 10$ Angstrom).

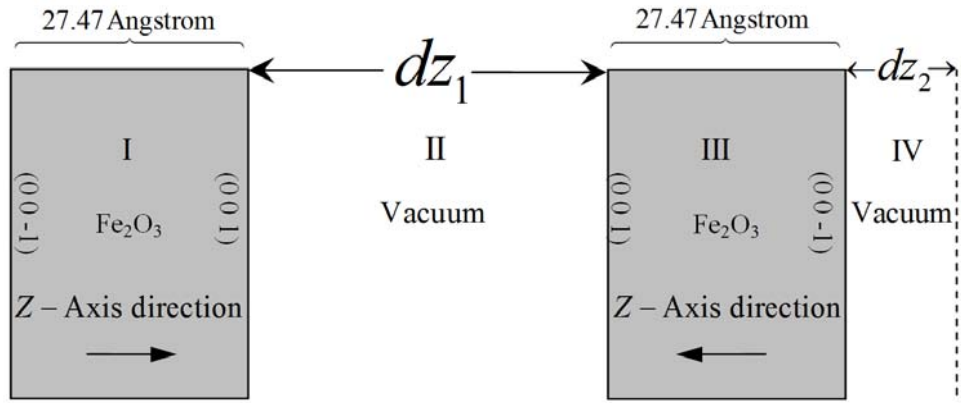

Fig. 2 Schematic illustration of the slab model.

## 3. Results and discussions

Figure 3 shows the electrostatic potential energy along the Z axis (averaged in XY plane). The Fermi level is set to zero. The constant potential energy at the middle of the vacuum slab has been checked. We obtained WFs 6.10 eV and 5.49 eV for (001) and (00-1) surface of $Fe_2O_3$ respectively.

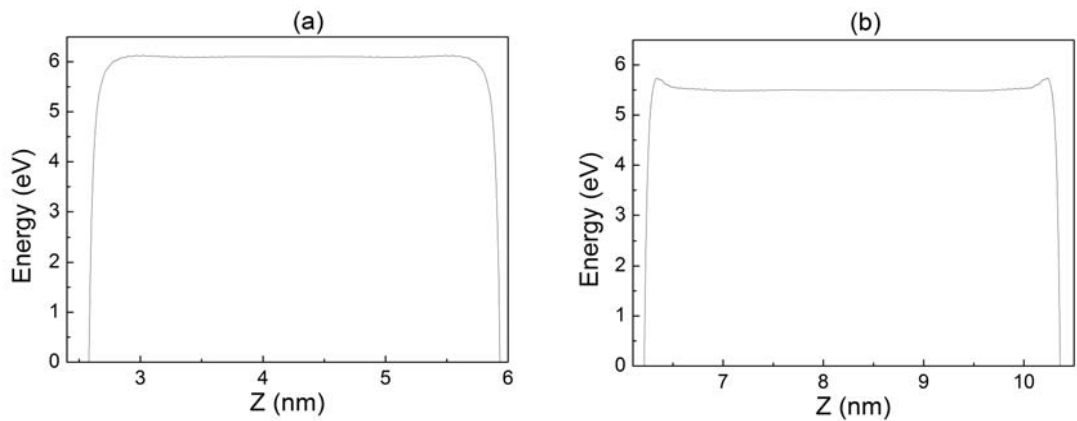

Fig. 3 Electrostatic potential curve in the Z-axis direction (averaged in XY plane). (a) is for (001) and (b) is for (00-1).

The charge density near the vacuum surface is very important to the vacuum potential energy. The atom which is closest to the surface is found, then we plot the charge density in the plane (001) which comes across this atom. Figure 4 (a) is for the case of surface (001) and (b) is for (00-1). The iron atoms are located in the center of the red contours and oxygen atoms correspond to the yellow ones. The color indicates the charge density: red corresponds to the maximum of charge density, and blue corresponds to the minimum of charge density. From figure 4, it is concluded that the WFs of these two surfaces are different because the charge density on these two surfaces is not the same (Fig. 4 (a) and (b)).

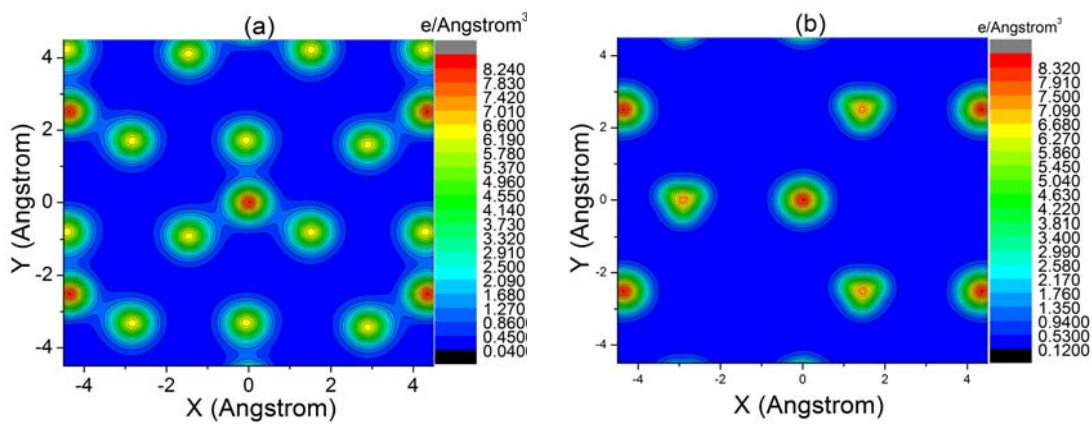

Fig. 4  Charge density distribution at (*a*) (001) and (*b*) (00 -1) surface of $Fe_2O_3$. The color indicates the charge density.

## 4. Conclusion

In summary, the atomic structure and the electron distribution of $Fe_2O_3$ are investigated with DFT. Symmetry slab model is used to eliminate the electric field caused by the polarized surfaces. The obtained WFs are 6.10 eV and 5.49 eV for the (001) and (00-1) surfaces respectively. For the pure iron, its WF is about 5.00eV. When the iron device is rust-eaten, the WF will be higher, so the drive power should be adjusted accordingly.

**ACKNOWLEDGMENTS**

The project is supported by the high-performance grid computing platform of Sun Yat-sen University.